\documentclass[iop]{emulateapj}
\usepackage{natbib}
\usepackage{graphicx}
\usepackage{hyperref}
\usepackage{breakurl}
\newcommand{\msun} {$M_{\odot}$}

\newcommand{\rsun} {$R_{\odot}$}

\newcommand{\hbeta} {H$\beta$}
\newcommand{\hgamma} {H$\gamma$}

\newcommand{\hepsilon} {H$\epsilon$}
\newcommand{\Te} {$T_{\rm eff}$}
\newcommand{\logg} {$\log g$}

\newcommand{\logca} {$\log$ (Ca/H)}
\newcommand{\logmg} {$\log$ (Mg/H)}
\newcommand{\logti} {$\log$ (Ti/H)}
\newcommand{\logcr} {$\log$ (Cr/H)}
\newcommand{\logfe} {$\log$ (Fe/H)}

\newcommand{\cai} {Ca {\sc i}}
\newcommand{\caii} {Ca {\sc ii}}
\newcommand{\mgi} {Mg {\sc i}}
\newcommand{\mgii} {Mg {\sc ii}}
\newcommand{\tii} {Ti {\sc i}}
\newcommand{\cri} {Cr {\sc i}}
\newcommand{\fei} {Fe {\sc i}}
\newcommand{\kms} {km s$^{-1}$}

\begin{document}

\slugcomment{\bf Accepted to ApJ}
\shortauthors{GIANNINAS ET AL.}
\shorttitle{J0745: DISCOVERY OF A METAL-RICH TIDALLY DISTORTED ELM WD}

\title{SDSS J074511.56+194926.5: DISCOVERY OF A METAL-RICH AND TIDALLY
  DISTORTED EXTREMELY LOW MASS WHITE DWARF}

\author{A.~Gianninas$^{1}$, J.~J.~Hermes$^{2,3}$, Warren~R.~Brown$^{4}
$, P.~Dufour$^{5}$, Sara~D.~Barber$^{1}$, Mukremin~Kilic$^{1}$, 
Scott~J.~Kenyon$^{4}$, and Samuel~T.~Harrold$^{2}$}

\affil{$^{1}$Homer L. Dodge Department of Physics and Astronomy,
  University of Oklahoma, 440~W.~Brooks~St., Norman, OK 73019, USA
  alexg@nhn.ou.edu}
\affil{$^{2}$Department of Astronomy, University of Texas at Austin,
  Austin, TX 78712, USA}
\affil{$^{3}$Department of Physics, University of Warwick, Coventry
  CV4 7AL, UK;}
\affil{$^{4}$Smithsonian Astrophysical Observatory, 60~Garden~St.,
  Cambridge, MA 02138, USA}
\affil{$^{5}$D\'epartement de Physique, Universit\'e de Montr\'eal,
  C.P.~6128, Succ.~Centre-Ville, Montr\'eal, Qu\'ebec H3C 3J7, Canada}

\begin{abstract}

We present the discovery of an unusual, tidally-distorted extremely
low mass white dwarf (WD) with nearly solar metallicity. Radial
velocity measurements confirm that this is a compact binary with an
orbital period of 2.6975~hrs and a velocity semi-amplitude of
$K$~=~108.7~\kms. Analysis of the hydrogen Balmer lines yields an
effective temperature of \Te~=~8380~K and a surface gravity of
\logg~=~6.21 that in turn indicate a mass of $M$~=~0.16~\msun\ and a
cooling age of 4.2 Gyr. In addition, a detailed analysis of the
observed metal lines yields abundances of \logmg~=~$-$3.90,
\logca~=~$-$5.80, \logti~=~$-$6.10, \logcr~=~$- $5.60, and
\logfe~=~$-$4.50, similar to the sun. We see no evidence of a debris
disk from which these metals would be accreted though the possibility
cannot entirely be ruled out. Other potential mechanisms to explain
the presence of heavy elements are discussed. Finally, we expect this
system to ultimately undergo unstable mass transfer and merge to form
a $\sim$~0.3--0.6~\msun\ WD in a few Gyr.

\end{abstract}

\keywords{binaries: close -- stars: abundances -- stars: atmospheres
  -- stars: individual (SDSS J074511.56+194926.5) -- techniques:
  photometric -- techniques: spectroscopic -- white dwarfs}

\section{INTRODUCTION}

Over the course of the last several years, we have successfully
targeted extremely low mass (ELM) white dwarfs (WDs) as part of the
ELM survey
\citep{brown_ELM1,brown_ELM3,brown_ELM5,kilic_ELM2,kilic_ELM4}.  These
WDs are characterized by surface gravities of 5.0~$<$~\logg~$<$~7.0
or, alternatively, with masses $M <$~0.25~\msun. ELM WDs are
necessarily the product of evolution within compact binary systems
since the Galaxy is not yet old enough to produce them through single
star evolution. This assertion is corroborated through radial velocity
measurements which confirm that all ELM WDs are in tight binary
systems with periods $P_{\rm orb} \leqslant$~1 day
\citep{brown_ELM5,kilic_ELM4}.  ELM WDs are understood to be He-core
WDs formed when the companion strips the outer envelope from a
post-main-sequence star before the star reaches the tip of the red
giant branch and ignites the helium.

ELM WD binaries afford us a unique opportunity to study the endpoint
of the evolution of compact binaries. Furthermore, these systems could
represent the eventual progenitors of Type Ia supernovae,
underluminous .Ia supernovae \citep{bildsten07}, and AM CVn systems
\citep{kilic13c}, as well as being strong sources of gravitational
waves that result in orbital decay \citep{hermes12b,kilic13a}.  Due to
the nature of these close binaries, they also afford us the
opportunity to study various phenomena which cause luminosity
variations in the light curves of these stars. These include
ellipsoidal variations due to tidal distortion, Doppler beaming, and
eclipses by the companion \citep{hermes12a,hermes12b}. The detection
and analysis of these effects is important because they provide
model-independent measurements of fundamental parameters such as the
stellar radius and mass.

Another point of interest is the fact that the optical spectra of all
ELM WDs with \logg\ $<$ 6.0 show the Ca K line at 3933 \AA. Indeed,
this phenomenon has been mentioned in most previous ELM papers but has
not yet been explored in a quantitative fashion. In higher mass WDs,
the presence of metallic absorption lines in their optical spectra is
explained by the ongoing accretion of material from a circumstellar
debris disk. These polluted WDs must be actively accreting material to
explain the observed abundances since diffusion timescales for heavy
elements in WD atmospheres are much shorter than their evolutionary
timescales. This paradigm is supported by the detection of debris
disks through infrared (IR) excess in the emission of more than two
dozen WDs \citep{jura03,kilic06,farihi09,barber12}. The heavy elements
observed in the spectra of otherwise pure hydrogen or helium
atmosphere WDs represent a means to study the composition of planetary
bodies that survived the late phases of stellar evolution
\citep{dufour10,klein10,klein11,zuckerman07}. There is growing
evidence that perhaps all metal-polluted WDs have acquired their heavy
material from an orbiting debris disk reservoir whose origin is
explained by the tidal disruption of one or many large rocky bodies
that ventured too close to the star
\citep{debes02,farihi10a,farihi10b,jura03,jura06,jura08,jura07,melis10}.
Such disks, which are easily detectable at IR wavelengths, are now
being detected at an accelerated pace with more than 20 cases
uncovered in the last 6 yr alone \citep[][and references therein]
{farihi09,farihi11,debes12,kilic12,hoard13}. Moreover, models which
aim to explore and understand the actual accretion of the debris disk
material onto the surface of the WD seem to be in good agreement with
observations \citep{rafikov11a,rafikov11b,rafikov12}.

On the other hand, recent calculations show that some metals (e.g., C,
Al, Si) can potentially be maintained in the atmospheres of WDs with
\logg~=~8.0 and as cool as \Te~=~20,000 K by radiative forces
\citep{chayer10}. If we consider that ELM WDs have surface gravities
two to three orders of magnitude weaker, radiative levitation may be
relevant for maintaining metals in the atmospheres of these WDs. What
is certain is that a detailed analysis of the metal abundances in ELM
WDs is a necessary first step in understanding the origin of the metal
pollutants in these stars. In this connection, the recent study of the
WD companion of PSR~J1816+4510 by \citet{kaplan13}, is most
intriguing. Their analysis yielded \Te\ = 16,000 $\pm$ 500 K and
\logg\ = 4.9 $\pm$ 0.3 with super-solar abundances of He, Mg, Si, Ca
and Fe.

In this paper, we present the detailed spectroscopic and photometric
analysis of SDSS J074511.56+194926.5 (hereafter J0745), a new
metal-rich and tidally distorted ELM WD in a compact binary. In
Section 2 we present our spectroscopic and photometric
observations. Next, Section 3 outlines the model atmospheres used in
our spectroscopic analysis and we present in Section 4 our analysis of
all the available spectroscopic and photometric data. This is
followed, in Section 5, by a discussion of the results and potential
evolutionary scenarios that would apply to J0745. Finally, we make
some concluding remarks and discuss future work in Section 6.

\begin{figure}
\includegraphics[scale=0.465,bb=35 167 692 640]{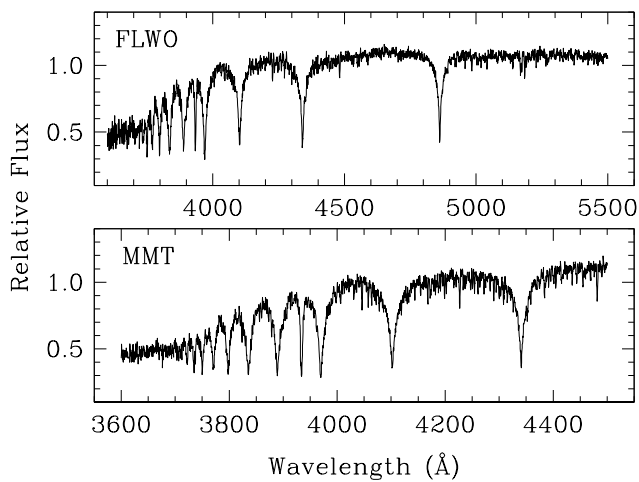}
\figcaption[f01.eps]{Radial-velocity corrected and summed optical
  spectra of J0745+1949. The spectra are normalized to unity at
  4200~\AA. The lower resolution spectrum taken at FLWO (top) covers a
  range from 3600 to 5500~\AA\ while the spectrum obtained at the MMT
  (bottom) covers a smaller spectral range from 3600 to 4500~\AA\ but
  at a higher resolution. \label{fg:spec}\\}
\end{figure}

\section{OBSERVATIONS}

\subsection{Optical Spectroscopy}

J0745 was first observed on 2010 November 5 using the FLWO 1.5~m
telescope and the FAST spectrograph \citep{fabricant98}.  We used the
600 line mm$^{-1}$ grating and a 1$\farcs$5 slit, providing wavelength
coverage from 3500~\AA\ to 5500~\AA\ and a spectral resolution of
1.7~\AA. Initial stellar atmosphere fits revealed J0745 to be a
candidate ELM WD. Follow-up spectra obtained one year later at the
FLWO 1.5~m revealed a significant change in radial velocity. We thus
obtained additional spectroscopy on five more nights to constrain
J0745's orbital solution. All spectra were paired with a comparison
lamp exposure, and have an average 13~\kms\ radial velocity
uncertainty.

We also obtained higher signal-to-noise (S/N) observations on 2012 May
23 with the 6.5~m MMT telescope and the Blue Channel spectrograph
\citep{schmidt89}.  We used the 832 line mm$^{-1}$ grating and a
1$\farcs$0 slit, providing wavelength coverage 3600~\AA\ to
4500~\AA\ and a spectral resolution of 1.0~\AA.  The individual MMT
spectra have 7~\kms\ radial velocity uncertainty; the summed MMT
spectrum has S/N~=~120 per resolution element. The summed FLWO and MMT
spectra are shown in Figure \ref{fg:spec}. Though the FLWO spectrum
provides better wavelength coverage, the MMT spectrum has a higher S/N
and a better resolution allowing for a much better identification of
all the observable metal lines. Consequently, the MMT spectrum will be
the basis of our atmospheric parameter and metal abundance
determinations.

\begin{table}\scriptsize
\caption{Available Broadband Photometry for J0745}
\begin{center}
\begin{tabular*}{\hsize}{@{\extracolsep{\fill}}lcr@{}}
\hline
\hline
\noalign{\smallskip}
Filter & $\lambda_{\rm eff}$ & Magnitude \\
       & ($\mu$m)            &           \\
\noalign{\smallskip}
\hline
\noalign{\smallskip}
$FUV$ & 0.153 & 22.321 $\pm$ 0.380 \\
$NUV$ & 0.229 & 18.665 $\pm$ 0.040 \\
\hline
$u$   & 0.355 & 17.426 $\pm$ 0.016 \\
      &       & 17.423 $\pm$ 0.016 \\
$g$   & 0.467 & 16.491 $\pm$ 0.008 \\
      &       & 16.505 $\pm$ 0.010 \\
$r$   & 0.617 & 16.488 $\pm$ 0.018 \\
      &       & 16.395 $\pm$ 0.011 \\
$i$   & 0.748 & 16.479 $\pm$ 0.011 \\
      &       & 16.458 $\pm$ 0.013 \\
$z$   & 0.893 & 16.564 $\pm$ 0.014 \\
      &       & 16.555 $\pm$ 0.020 \\
\hline
$J$   & 1.241 & 15.915 $\pm$ 0.058 \\
$H$   & 1.651 & 16.156 $\pm$ 0.171 \\
$K$   & 2.166 & 15.655 $\pm$ 0.172 \\
\hline
$W1$  & 3.379 & 15.819 $\pm$ 0.068 \\
\noalign{\smallskip}
\hline
\label{tab1}
\end{tabular*}
\end{center}
\end{table}

\subsection{Time-series Photometry}

We obtained a total of 6.6 hr of time-series photometry for J0745
using the Argos frame transfer camera mounted on the McDonald 2.1~m
telescope \citep{nather04}. Observations were obtained over the course
of two observing runs in 2012~November (4.9 hr) and 2013~March (1.7
hr). Our exposures ranged from 10 to 15 s with negligible read-out
time, and were obtained through a 3 mm {\em BG40} filter to reduce sky
noise.

We performed weighted, circular, aperture photometry on the calibrated
frames using the external IRAF package $\textit{ccd\_hsp}$ written by
Antonio Kanaan \citep{kanaan02}. We divided the sky-subtracted light
curves by the brightest comparison stars in the field to correct for
transparency variations, and applied a timing correction to each
observation to account for the motion of the Earth around the
barycenter of the solar system \citep{stumpff80,thompson09}. We have
also reduced, in an identical manner, the bright nearby star
SDSS~J074508.34+194958.4. This star has a similar $g-r$ color as our
target, and helps us constrain the differential atmospheric effects on
our time-series photometry.

\subsection{Broadband Photometry}

In Table \ref{tab1}, we list all of the available broadband photometry
for J0745 from ultraviolet (UV), optical and IR surveys. There are two
available sets of $ugriz$ magnitudes from the Sloan Digital Sky Survey
(SDSS) Data Release 9 \citep{ahn12}. The first set, corresponding to
the first of each pair of $ugriz$ magnitudes listed in Table
\ref{tab1}, was obtained on 2003 January 25 while the second
observations were taken a week later on 2003 February 1. J0745 was
also detected in the UV by the {\it Galaxy Evolution Explorer}
\citep[$GALEX$,][]{martin05}. In addition, IR photometry is available
from the Two Micron All Sky Survey \citep[2MASS;] []{skrutskie06} and
the {\it Wide-field Infrared Survey Explorer}
\citep[$WISE$;][]{wright10}. We only list the $WISE$ magnitude in the
$W1$ band since the S/N of the $W2$, $W3$ and $W4$ band observations
is S/N $<$ 4.

\section{MODEL ATMOSPHERES}

\subsection{Pure Hydrogen Model Atmospheres}

We use the hydrogen-rich model atmospheres described at length in
\citet{tremblay10}, and references therein, for the analysis of the
hydrogen Balmer lines, from which we obtain our measurements of
\Te\ and \logg. Briefly, these models assume a plane-parallel
geometry, hydrostatic equilibrium and local thermodynamic equilibrium
(LTE). The assumption of LTE is justified as our model grid is
restricted to \Te\ $<$ 30,000~K where NLTE effects are not yet
significant. Furthermore, these models include the new Stark
broadening profiles from \citet{TB09} that include the occupation
probability formalism of \citet{hm88} directly in the line profile
calculation. Our grid covers \Te\ from 4000 K to 30,000~K in steps
ranging from 250 to 5000~K, and \logg\ from 5.0 to 9.5 in steps of
0.25 dex.

\subsection{Model Atmospheres with Metals}

We calculated metal-blanketed model atmospheres and synthetic spectra
using the same code that was used to model the heavily metal polluted
DBZ star J0738+1835 \citep{dufour12}. As a first step, we calculated a
pure hydrogen model using the atmospheric parameters (\Te\ and \logg)
as determined from the fits to the Balmer lines. We then compared this
model with an equivalent model computed using the code described in
the previous section and confirmed that the thermodynamic structures
were identical. Next, we proceeded to calculate several grids of
synthetic spectra, one for each element of interest (i.e. Mg, Ca, Ti,
Cr, and Fe), keeping the abundance of all other elements fixed to
their expected values \citep[by assuming CI chondrite ratios
  from][]{lodders03}.  The individual grids cover a range of
abundances from $\log[n(\rm Z)/n(\rm H)]$~=~$-$3.0 to $-$10.0, in
steps of 0.5~dex. These grids are then used to determine the different
elemental abundances.

\begin{figure}
\includegraphics[scale=0.5,bb=65 167 692 604]{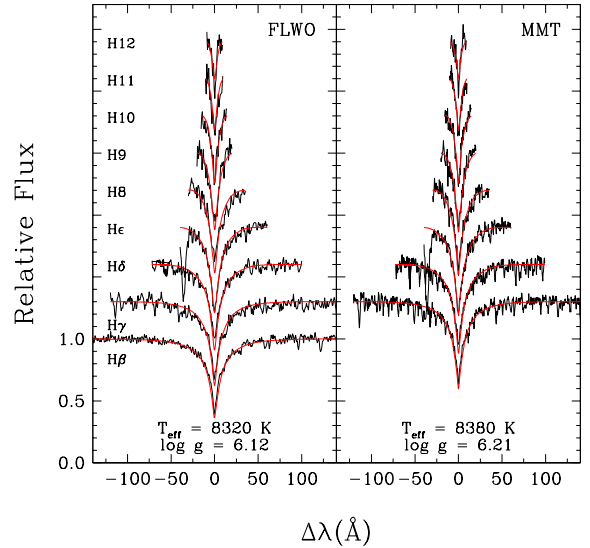}
\figcaption[f02.eps]{Model atmosphere fits (red) to the observed
  hydrogen Balmer lines (black) from \hbeta\ to H12 for the spectrum
  taken at FLWO (left) and from \hgamma\ to H12 for the spectrum
  obtained at the MMT (right). The values of \Te\ and \logg\ obtained
  from the individual fits are given at the bottom of each panel. The
  individual Balmer lines are normalized to unity and offset
  vertically by a factor of 0.3 for clarity. Note that we exclude the
  spectral range containing the Ca K line, from 3925 to 3945 \AA\,
  from both the normalization procedure and the fitting
  routine. \label{fg:linefit}}
\end{figure}

\begin{figure*}
\includegraphics[scale=0.7,angle=-90,bb=62 15 596 764]{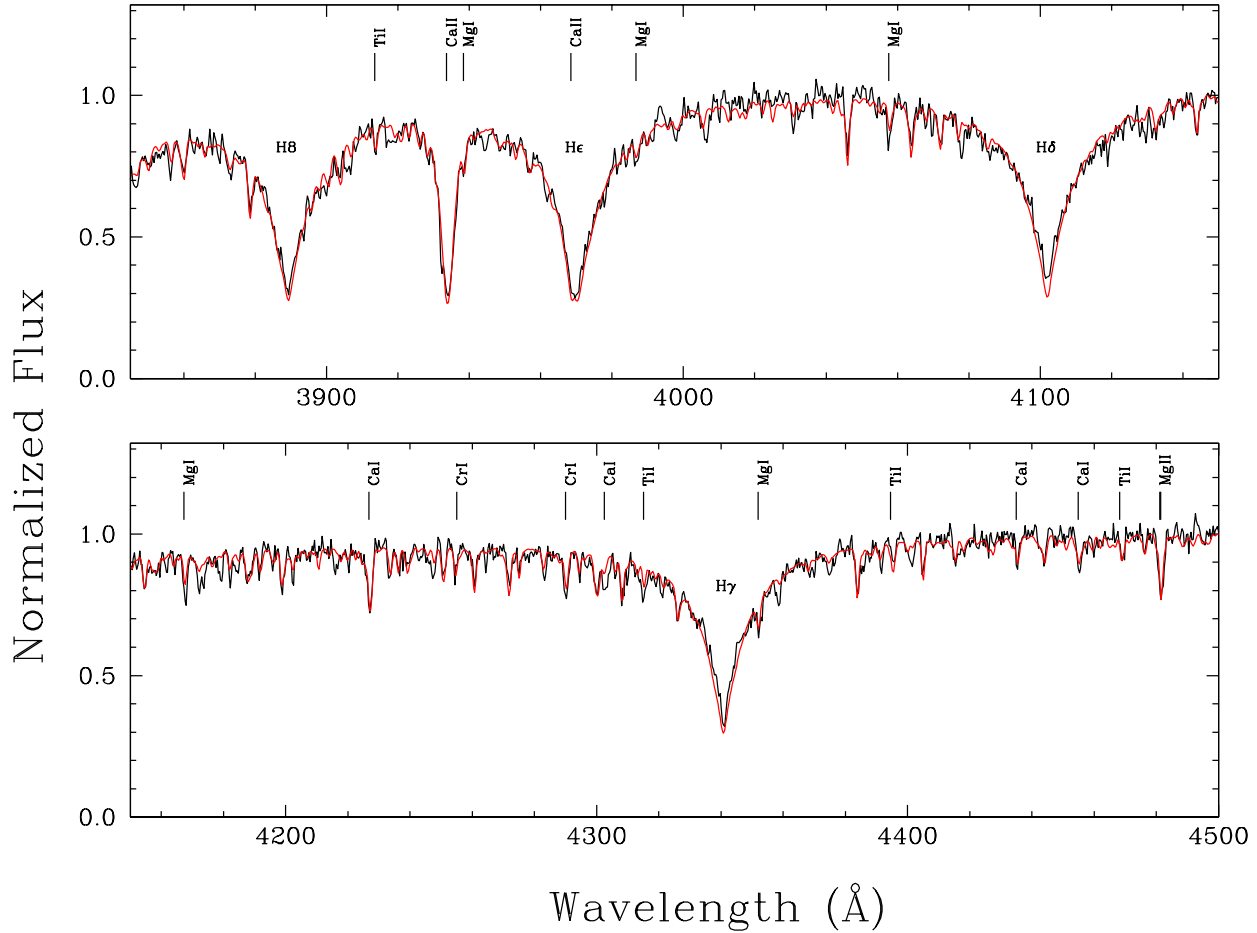}
\figcaption[f03.eps]{Our best fit metal blanketed model (red) using
  \Te\ and \logg\ from the Balmer line fits plotted over the observed
  MMT spectrum (black) of J0745. The Balmer lines from \hgamma\ to H8
  are indicated in the figure as well as lines of \mgi, \mgii, \cai,
  \caii, \tii, and \cri. All the remaining lines are
  \fei\ lines. \label{fg:metals}}
\end{figure*}

\section{ANALYSIS}

\subsection{Atmospheric Parameters}

Our Balmer line fits use the so-called spectroscopic technique
described in \citet{gianninas11} and references therein. One important
difference between our procedure and that of \citet{gianninas11} is
that we fit higher-order Balmer lines, up to and including H12,
observed in low surface gravity ELM WDs. The higher-order Balmer lines
are quite sensitive to \logg\ and consequently improve our surface
gravity measurement. Our error estimates combine the internal error of
the model fits, obtained from the covariance matrix of the fitting
algorithm, and the external error, obtained from multiple observations
of the same object. External uncertainties are typically 1.2\% in
\Te\ and 0.038 dex in \logg\ \citep[see][for details]{LBH05}.

We present in Figure \ref{fg:linefit} our fits to the normalized
Balmer lines for both the FLWO (left) and MMT (right) spectra. We
exclude the wavelength region from 3925~\AA\ to 3945~\AA, in the blue
wing of \hepsilon, from both our normalization and fitting procedures
due to the presence of the strong \caii\ K line at 3933~\AA. The fits
yield \Te~$=8320\pm120$~K and \logg~$=6.12\pm0.09$ for the FLWO
spectrum and \Te~$=8380\pm120$~K and \logg~$=6.21\pm0.07$ for the MMT
spectrum. The somewhat large uncertainties in \logg\ are likely due to
the presence of the many metal lines in the wings of the Balmer lines
which makes it difficult to accurately model the line widths which are
very sensitive to variations in \logg. Nonetheless, the results are
consistent within the uncertainties and we adopt the atmospheric
parameters obtained from the MMT spectrum for the remainder of our
analysis owing to the higher S/N of those observations.

In order to determine a mass for J0745, we employ the new evolutionary
calculations for ELM WDs presented in \citet{althaus13}. In fact, we
have employed a slightly finer grid of models than the ones published
in the aforementioned article, kindly provided to us by L. Althaus
(2013, private communication). These models predict a mass of
$M_{1}=0.164\pm0.003$~\msun\ as well as a cooling age of $\tau_{\rm
  cool}=4.2\pm0.6$~Gyr. The uncertainty in mass (and age) is a
combination of the uncertainties provided for each model in the grid
with our uncertainties in \Te\ and \logg. However, we suspect that the
model uncertainties are somewhat optimistic and we adopt an
uncertainty of 0.16~$\pm$~0.02~\msun\ ($\sim$~10\%).

Given the mass, we then combine \logg\ and $M_{1}$ to compute a radius
of $R_{1}$~=~0.053~$\pm$~0.008~\rsun. Next, by coupling our
determinations of \Te, \logg, and $R_{1}$ with the photometric
calibrations of \citet{holberg06} we obtain an absolute magnitude of
$M_{g}$~=~9.8~$\pm$~0.4. Finally, we are able to combine the absolute
and apparent magnitudes of J0745 to calculate a distance of
$d$~=~201~$\pm$~36~pc. These derived quantities should be viewed with
some caution considering the uncertainties in the evolutionary models
on which they depend. Future observations and improved model
calculations will permit more robust determinations. In particular,
the distance to J0745 can be confirmed through ground- or space-based
parallax observations such as will become available with the upcoming
{\it Gaia} mission \citep{perryman01}.

\subsection{Metal Abundances}

Our method for fitting the multitude of metal lines in the spectrum of
J0745 follows the procedure outlined in \citet{dufour12} which
considers 15--30~\AA\ wide segments centered on each of the strongest
absorption features rather than fitting each individual line. We note
that in many cases, more than one line from a given fitted element can
be found in a single segment. In these cases, the measured abundance
from that segment is essentially a weighted average of the lines, more
weight (more frequency points) being attributed to the strongest
lines. We finally take as the measured abundance of an element the
average over all the segments.

We display in Figure \ref{fg:metals} our best fit solution for J0745
which includes Mg, Ca, Ti, Cr, and Fe. We determine abundances, by
number, of \logmg~$=-3.90\pm0.30$, \logca~$=-5.80\pm0.30$, \logti~
$=-6.10\pm0.30$, \logcr~$=-5.60\pm0.30$, and
\logfe~$=-4.50\pm0.30$. The uncertainties on the abundances reflect
the observed dispersion in the abundance measurements obtained from
the individually fitted segments. The large dispersion can be
attributed to a combination of the spectral resolution as well as the
strength (near the noise level in many cases) and total number of
observed metal lines. A high-S/N, high-resolution spectrum will permit
more accurate determinations.

We see in Figure \ref{fg:metals} that the observed line strengths are
not all reproduced exactly though the overall match with our
spectroscopic solution is quite good, particularly for most of the
strong Ca and Mg features in the spectrum. Moreover, we remark that
the abundances for Mg, Ti, and Cr are super-solar, the Ca abundance is
slightly sub-solar and the Fe abundance is approximately solar when
compared to the solar abundances listed in \citet{asplund05}.

Finally, the fact that these abundances were determined from synthetic
spectra calculated from a structure corresponding to an atmosphere of
pure hydrogen could be considered as an additional source of
uncertainty. It is true that in hot WDs, metals can significantly
affect the thermodynamic structure of the atmosphere since ionized
metals are important sources of opacity. However, J0745 is too cool
for the metals to have a similar impact. In their analysis of GD 362
(which is $\sim$1500~K hotter than J0745), \citet{gianninas04} noted
that synthetic spectra computed from a hydrogen pure model atmosphere
and a metal-blanketed model were identical. However, our solution
cannot be considered final since our analysis was done for fixed
values of \Te\ and \logg\ as measured from the Balmer line fits. The
relative composition of the polluting elements, however, is much less
sensitive to the exact final parameters adopted and can thus be used
as a high-precision measurement of the accreted material if radiative
levitation is negligible \citep[see][]{chayer13}.

\begin{figure}
\includegraphics[scale=0.435,bb=10 125 492 709]{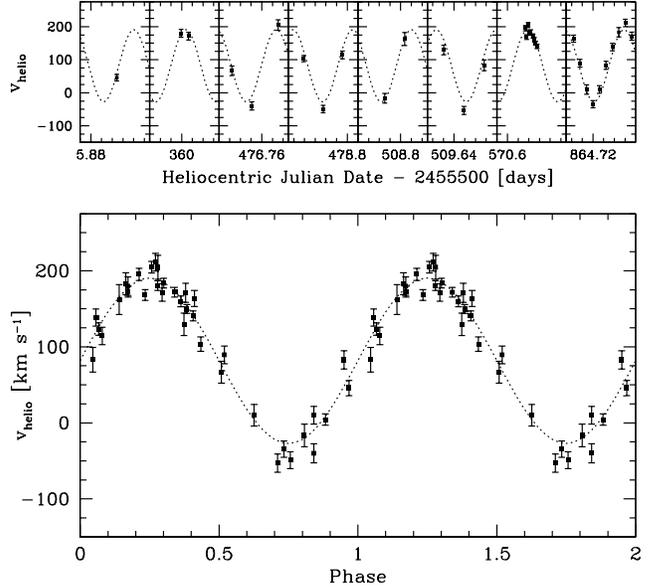}
\figcaption[f04.eps]{Radial velocities vs.\ time, plotted with a tick
  size of 0.02 days (upper panel), and phased to the best-fit
  spectroscopic orbital solution (lower panel). \label{fg:rvplot}}
\end{figure}

\subsection{Orbital Solution}

Figure \ref{fg:rvplot} plots the radial velocities versus time (upper
panel) and the radial velocities phased to the best-fit orbital
solution (lower panel). We calculate orbital elements using the code
of \citet{kenyon86} as described in our earlier ELM Survey papers
\citep[][and references therein]{brown_ELM5}. J0745 has a velocity
semi-amplitude $K=108.7\pm2.9$ \kms, a systemic velocity
$\gamma=81.7\pm4.3$ \kms, and an orbital period $P_{\rm spec}
=2.6975\pm0.2$ hr. The uncertainty in radial velocity orbital period
is large because there are period aliases near the best-fit value, at
2.7078, 2.6873, and 2.4326 hr. Based on these determinations, the
resultant binary mass function is $f=0.015\pm0.002~M_{\odot}$.

For an edge-on orbit ($i=90^{\circ}$), the mass function provides a
lower limit for the companion mass which in this case yields $M_{2} >
0.102\pm0.014$~\msun. If we assume the mean inclination angle for a
random stellar sample, $i=60^{\circ}$, we get an estimate of the most
probable companion mass. For J0745, the most likely companion mass is
$0.124\pm0.017$~\msun. This companion mass is still rather low, even
by ELM WD standards, and so we suspect that this is a low-inclination
system with an older, fainter, and likely more massive WD companion
(see Section \ref{sc:phot} for further discussion). Alternatively,
assuming a random orbital inclination distribution allows us to
calculate the probability of different companion masses. Specifically,
there is a 92\% probability that the companion is another low-mass WD
with $M_{2} \leqslant 0.45$~\msun.

Finally, general relativity predicts that short period binaries like
J0745 lose orbital angular momentum due to gravitational wave
radiation. The timescale for the binary to shrink and begin mass
transfer via Roche-lobe overflow is $\tau_{\rm merger}\leqslant
5.4$~Gyr. If we assume the most probable companion mass of 0.12~\msun,
J0745 will begin mass transfer in $<4.6$~Gyr. However, if J0745 really
is a low-inclination system, then it will begin merging in only a few
Gyr (e.g., $\tau_{\rm merger}\sim 2$~Gyr for $i=30^{\circ}$).

\begin{figure}
\includegraphics[scale=0.5,bb=110 177 596 705]{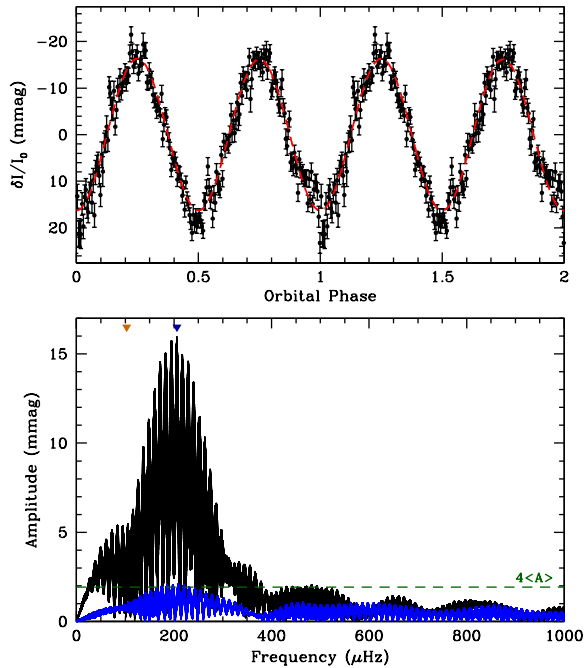}
\figcaption[f05.eps]{Time-series photometry of J0745+1949. The top
  panel shows the optical light curve, folded at the orbital period
  and repeated for clarity. The dashed red line represents the
  best-fit orbital solution. The bottom panel shows a Fourier
  transform of the target (black) and brightest comparison star
  (blue). The orange and dark blue triangles mark the orbital period
  and half-orbital period, respectively, as determined from the
  time-series spectroscopy. Variability in the comparison star shows
  evidence of atmospheric variability at similar timescales as the
  period of interest in the target, which may explain why the
  photometry does not phase exceptionally well. The spectroscopy and
  photometry both find $\sim$~2.7 hr as the best-fit orbital
  period. \label{fg:lc}}
\end{figure}

\subsection{Light Curve Modeling}

Figure \ref{fg:lc} shows the Argos light curve for J0745 (top panel)
folded over the best-fit orbital period. The dashed red line
corresponds to the best-fit orbital solution, as discussed in the
following analysis. Additionally, we have computed a Fourier transform
of our time-series photometry (Figure \ref{fg:lc}, bottom panel). With
less than 7 hr of data spread over five months, there is considerable
aliasing around our highest peak, which occurs at $205.850\pm0.063$
$\mu$Hz. Assuming this signal is the result of ellipsoidal variations
peaking twice per orbit, this corresponds to an orbital period of
$P_{\rm orb}=2.69883\pm0.00082$ hr. We use this period to fold our
time-series photometry in the top panel; this period agrees well with
the orbit derived from the spectroscopy of 2.6975 hr.

\begin{figure}
\includegraphics[scale=0.4,bb=0 145 596 644]{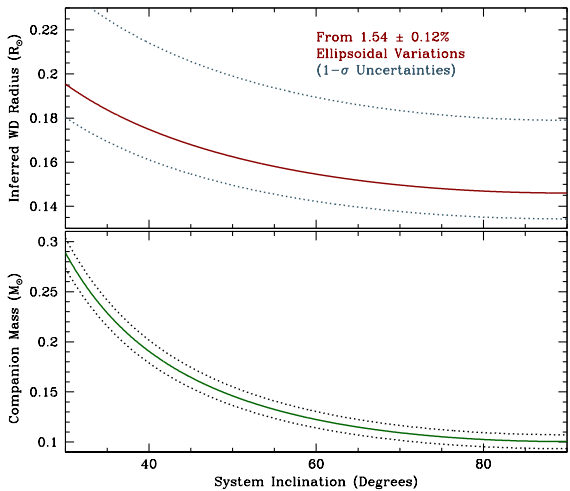}
\figcaption[f06.eps]{Top panel shows allowed values for the radius,
  modulo the unknown system inclination, to explain the 1.54\% $\pm$
  0.12\% amplitude ellipsoidal variations observed in J0745+1949, and
  includes 1$\sigma$ uncertainties. The larger upper-limit is a result
  of differing the gravity darkening coefficient. If we use the
  determined surface gravity and mass to predict an expected ELM WD
  radius, we do not expect an ELM WD with a radius larger than 0.06
  \rsun, which is much smaller than the lower-limit we measure from
  the ellipsoidal variations. The bottom panel shows the corresponding
  companion mass from the spectroscopic mass function, including
  1$\sigma$ uncertainties from the derived primary
  mass.  \label{fg:radius}}
\end{figure}

There are five major effects that can cause photometric variability in
a binary system: Doppler beaming, reprocessed light (reflection),
ellipsoidal variations, eclipses, and pulsations. To best characterize
the variability in the light curve, we perform a Monte Carlo analysis
by creating $10^5$ synthetic light curves in which we replace the
measured flux $f$ with $f + g~\delta f$, where $\delta f$ is the error
in flux and $g$ is a Gaussian deviate with zero mean and unit variance
(e.g. \citealt{hermes12a}). We then fit each light curve with a
five-parameter model that includes an offset, and the sine and cosine
terms that represent Doppler beaming ($\sin \phi$), ellipsoidal
variations ($\cos 2\phi$), reflection ($\cos \phi$), and the first
harmonic of the orbital period ($\sin 2\phi$).

We detect a high-amplitude signal consistent with ellipsoidal
variations of the low-mass, visible primary; this is manifest as a
$\cos 2\phi$ modulation in the light curve. Our Monte Carlo analysis
yields an amplitude of $\cos 2\phi = 1.54\pm0.12$\% for the
ellipsoidal variations; no other (co)sine terms appear with any
statistical significance. Although we expect a 0.15\% Doppler beaming
signal given $K_1$, the temperature of the primary, and our
blue-bandpass filter \citep{shporer10}, we do not detect this
low-amplitude $\sin \phi$ modulation at the orbital period with any
significance.

Tidal distortions do not cause a perfectly ellipsoidal shape, so we
represent the ellipsoidal variations as harmonics to the first four
$\cos \phi$ terms as derived in \citet{morris93}. Equation~(1) in that
work theoretically predicts the ellipsoidal variation amplitude, which
is dominated by

\begin{equation}
L(\phi)/L_{0}=\frac{-3(15-u_{1})(1+\tau_{1})(R_{1}/a)^{3}q\sin^{2}i}{20(3-u_{1})}\cos 2\phi
\end{equation}

\noindent where $q=M_{2}/M_{1}$ is the mass ratio, $u_{1}$ and
$\tau_{1}$ are the linear limb-darkening and gravity-darkening
coefficients for the primary, respectively, $a$ is the orbital
semi-major axis of the system, $i$ is the system inclination and
$R_{1}$ is the radius of the primary. Here we will assume $u_{1}$=
0.58 for this relatively cool ELM WD, as calculated by
\citet{gianninas13}.

We use the 1.54\%~$\pm$~0.12\% amplitude ellipsoidal variations to
infer the radius of the ELM WD primary as a function of the
inclination angle, since all unknown parameters in Equation~(1) can be
written in terms of $i$. This result is shown in
Figure~\ref{fg:radius}. We assume that energy transport at the surface
is primarily radiative and thus the gravity darkening coefficient is
$\tau_1 = 1.0$, but we have also considered $\tau_1 = 0.36$ for the
non-radiative case in our 1-$\sigma$ uncertainties shown in
Figure~\ref{fg:radius}.

The radius inferred from the ellipsoidal variations of the primary
appears considerably larger than we would expect given its surface
gravity and mass; we expect a radius of 0.053~\rsun\ from the
\citet{althaus13} models, but divergently we observe a minimum radius
at high inclinations of more than 0.13 \rsun. We caution that system
parameters derived from the ellipsoidal variation amplitude using the
approximation of \citet{morris93} are only valid when the primary is
corotating with the binary orbit, as was demonstrated for KOI-74 by
\citet{bloemen12} and may not be the case here. We discuss this
discrepancy further in Section~\ref{gravity}.

The low \logg\ and \Te\ (8380~K) of J0745 put it near the instability
strip for pulsations in ELM WDs \citep{hermes13}. However, we see no
evidence for variability at timescales other than the orbital- and
half-orbital periods, to a limit of 0.4\% in relative amplitude.

\begin{figure*}
\includegraphics[scale=0.685,angle=-90,bb=122 6 496 784]{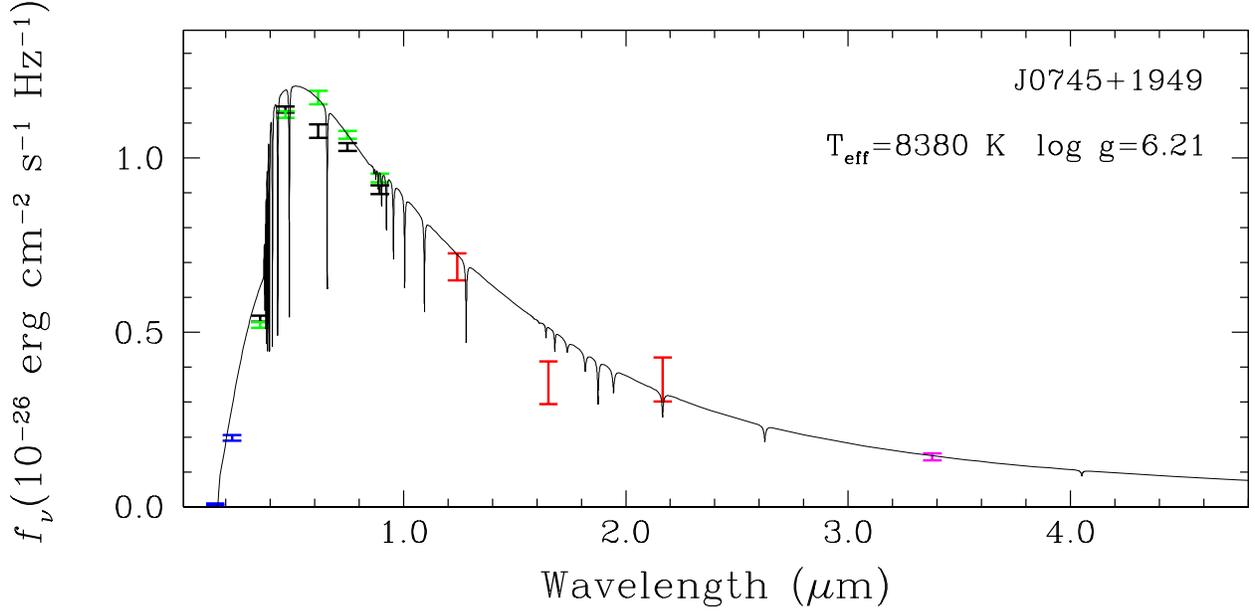}
\figcaption[f07.eps]{Plot of the combined $GALEX$ (blue), SDSS (2013
  January, black; 2013 February, green), 2MASS (red) and $WISE$
  (magenta) photometry. The black line represents the spectral energy
  distribution for our best-fit spectroscopic solution scaled to the
  observed photometry by our radius and distance determinations (see
  Equation (\ref{eq:phot})). \label{fg:phot}}
\end{figure*}

\subsection{Spectral Energy Distribution}\label{sc:phot}

The available UV-optical-IR photometry for J0745 is listed in Table
\ref{tab1}. Since J0745 is situated at a galactic latitude of $b
\approx 20^{\circ}$, interstellar extinction must be taken into
consideration. Consequently, we have applied corrections to both the
$GALEX$ and SDSS photometry. For the $GALEX$ bands we have adopted
extinction coefficients of $A_{\rm FUV}/E(B-V) = 8.24$ and $A_{\rm
  NUV}/(E(B-V)$ = 8.2, as prescribed by \citet{wyder07}, and combined
these with a reddening of $E(B-V) = 0.0614$ as provided for the
coordinates of
J0745.\footnote{\burl{http://galex.stsci.edu/GR6/?page=explore\&photo=true\&objid=6407287600137114823}}
In addition, we adopt the extinction corrections for each SDSS band as
provided on the SDSS
Skyserver.\footnote{\burl{http://skyserver.sdss3.org/dr9/en/tools/explore/obj.asp?ra=116.29817\&dec=19.82403}}
Both the $GALEX$ and SDSS extinction corrections are based on the dust
maps of \citet{schlegel98}. The corrected photometry is displayed in
Figure \ref{fg:phot} as the colored error bars. We plot as well the
full spectral energy distribution (SED) for a model corresponding to
our adopted spectroscopic solution of \Te~=~8380~K and
\logg~=~6.21. Furthermore, we have scaled the flux of the SED using
our determinations of the radius and distance (see Table \ref{tab2})
as the observed flux, $f_{\lambda,m}$, and the Eddington flux,
$H_{\lambda,m}$, are related by,

\begin{equation}
f_{\lambda,m} = 4\pi(R/d)^{2}H_{\lambda,m}
\label{eq:phot}
\end{equation}

\noindent where $R/d$ is the ratio of the radius of the star to its
distance from Earth. The agreement between the SED of our
spectroscopic solution and the observed photometry is quite good
overall. We do note a significant difference in the two sets of SDSS
photometry, particularly in the $r$-band measurements. Some of this
variation can be attributed to the observed photometric variability of
J0745 due to ellipsoidal variations. Furthermore, we do not see any
features that would signal the presence of the companion. If we assume
that the companion is indeed a more massive and cooler WD then it is
necessarily much fainter. Finally, there is no evidence of any IR
excess from a potential debris disk from which J0745 would actively be
accreting the metals observed in its optical spectrum.

\begin{table}\scriptsize
\caption{System Properties for J0745}
\begin{center}
\begin{tabular*}{\hsize}{@{}l@{\extracolsep{\fill}}r@{\extracolsep{0pt}}@{ $\pm$ }l@{}}
\hline
\hline
\noalign{\smallskip}
\Te (K)               &    8380 & 120    \\
\logg                 &    6.21 & 0.07   \\
\logmg                & $-$3.90 & 0.30   \\
\logca                & $-$5.80 & 0.30   \\
\logti                & $-$6.10 & 0.30   \\
\logcr                & $-$5.60 & 0.30   \\
\logfe                & $-$4.50 & 0.30   \\
$P_{\rm spec}$ (hr)   &  2.6975 & 0.2    \\
$P_{\rm phot}$ (hr)   &  2.6988 & 0.0008    \\
$K$ (\kms)            &   108.7 & 2.9    \\
$\gamma$ (\kms)       &    81.7 & 4.3    \\
$M_g$                 &    9.75 & 0.39   \\
$d$ (pc)              &     201 & 36     \\
$M_{1}$ (\msun)       &    0.16 & 0.02   \\
$R_{\rm 1, spec}$ (\rsun)       &   0.053 & 0.008  \\
$\tau_{\rm cool}$ (Gyr)   &   4.232 & 0.594  \\
Mass function (\msun) &   0.015 & 0.002  \\
$M_{2}$ (\msun)       & $>$0.102 & 0.014 \\
$M_{2,i=60^{\circ}}$ (\msun) & 0.124 & 0.017 \\
$\tau_{\rm merger}$ (Gyr) &$<$5.447 & 1.374 \\
$a$ (AU)              & $>$0.0029 & 0.0003 \\            
$a$ (\rsun)           & $>$0.630 & 0.058 \\
\noalign{\smallskip}
\hline
\label{tab2}
\end{tabular*}
\end{center}
\end{table}

\section{DISCUSSION}

\subsection{Origin of Metals}

\begin{figure}
\includegraphics[scale=0.45,bb=32 66 496 604]{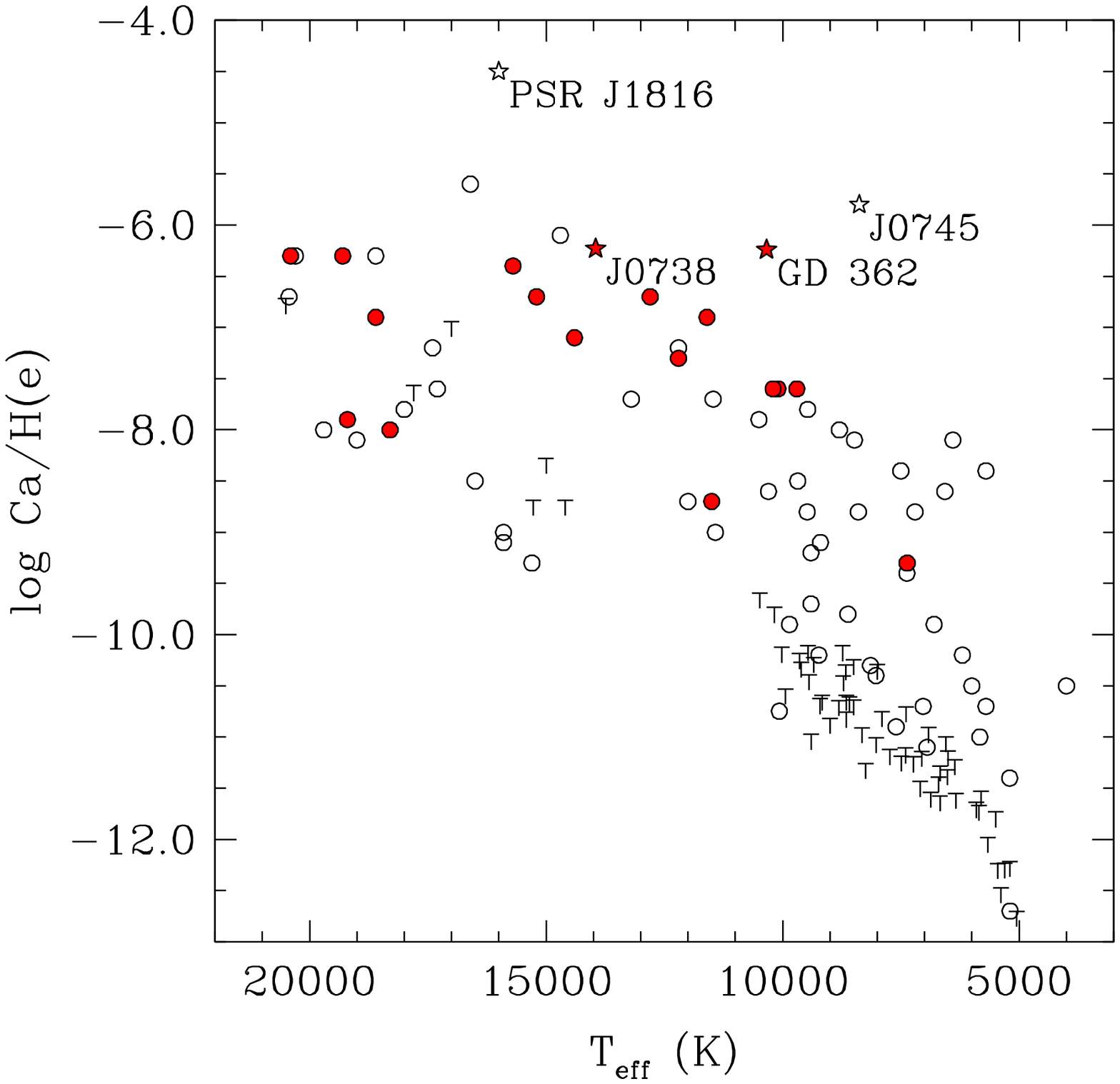}
\figcaption[f08.eps]{Published Ca abundances (white circles), or upper
  limits (crosses), for 133 polluted WDs given with respect to the
  main atmospheric constituent (i.e. \logca\ or $\log $~(Ca/He)), as a
  function of \Te. Abundance determinations are taken from
  \citet{zuckerman03,zuckerman07}, \citet{berger05},
  \citet{koester05}, \citet{farihi09,farihi10b}, \citet{dufour12}, and
  \citet{kaplan13}.  Stars denote objects of particular interest which
  are discussed in the text. Finally, filled red symbols correspond to
  WDs with known debris disks. \label{fg:Z_t}}
\end{figure}

\begin{figure}
\includegraphics[scale=0.45,bb=32 66 496 604]{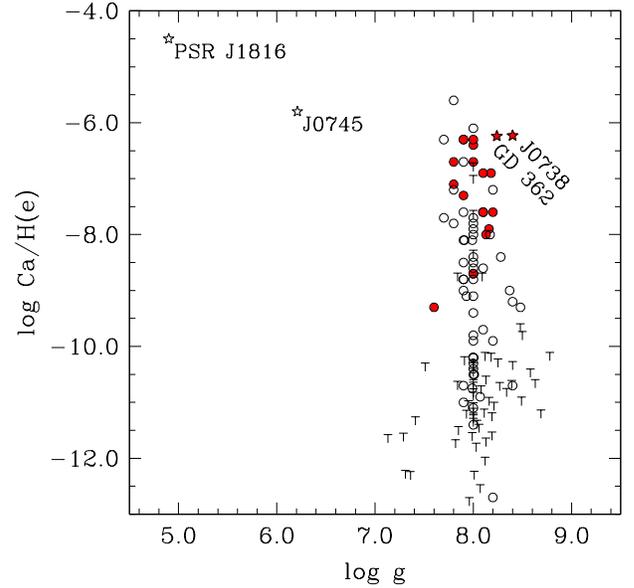}
\figcaption[f09.eps]{Same as Figure \ref{fg:Z_t} but with the Ca
  abundances plotted as a function of \logg\ instead. \label{fg:Z_g}}
\end{figure}

J0745 is truly a unique system that allows us to study several
phenomena simultaneously. First and foremost are the very high metal
abundances we observe. Indeed, J0745 stands out when compared to other
metal polluted WDs. In Figure \ref{fg:Z_t} we plot published Ca
abundances (or upper limits, see figure caption) for metal-polluted
WDs as a function of \Te. The WD companion of PSR~J1816+4510
\citep{kaplan13} is also shown and currently has the highest Ca
abundance measured for a WD. Figure \ref{fg:Z_t} also includes the
extremely metal-rich DBZ, J0738+1835 \citep{dufour12}, and GD~362
\citep{zuckerman07}. J0745 clearly stands apart from the majority of
the other WDs in Figure \ref{fg:Z_t} and has the highest Ca abundance
when compared to WDs with similar \Te. Similarly, Figure \ref{fg:Z_g}
shows the same Ca abundances plotted as a function of \logg. Most
objects cluster around \logg~$\sim$~8.0 while GD~362 and J0738 are
somewhat more massive. On the other hand J0745 and the WD companion to
PSR J1816 have \logg\ roughly two and three orders of magnitude lower,
respectively, highlighting once again the extreme nature of these
objects. Naturally, this leads us to inquire about the source of these
metals.

\begin{figure}
\includegraphics[scale=0.41,angle=-90,bb=122 31 496 584]{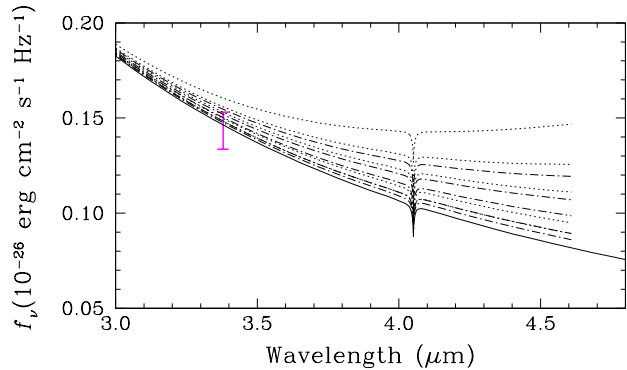}
\figcaption[f10.eps]{Similar to Figure \ref{fg:phot} but showing only
  the wavelength region with $\lambda > 3.0$~$\mu$m. The dotted lines
  represent the flux from disk models with outer disk radii of $R_{\rm
    out}$~=~1.4, 1.5, 2.0, 3.0, 4.0, 5.0~\rsun, from bottom to top,
  respectively, and for an inclination angle of $i=30^{\circ}$. The
  dash-dotted lines represent equivalent disk models but assuming,
  instead, an inclination angle of $i=60^{\circ}$. All the disk models
  assume an inner disk radius of $R_{\rm in} =
  1.3$~\rsun. \label{fg:disk}}
\end{figure}

The currently accepted paradigm for explaining the presence of metals
in more massive cool WDs is accretion from a circumstellar debris
disk. Indeed, we see in Figure \ref{fg:Z_t} that debris disks around
many of the polluted WDs have been detected. However, we have seen
that there is no readily observable IR excess in the available IR
photometry of J0745. Debris disks around more massive WDs tend to be
found within a $\sim$0.1--1.0~\rsun\, the typical tidal disruption
radius for WDs \citep{vh07,farihi10b}. However, we expect that in the
case of ELM WD binaries these debris disks will be circumbinary and
not simply circumstellar. Combining our determinations for the mass of
the primary $M_{1}$, the minimum companion mass, $M_{2}$, and the
orbital period, $P_{\rm spec}$, we compute a minimum orbital
separation of $a>0.0029\pm0.0003$~AU or equivalently
$a>0.63\pm0.06$~\rsun. In order to explore whether a debris disk can
explain the observed heavy elements, we have computed a series of
debris disk models following the formalism of \citet{jura03}. We note
that our models assume an optically thick disk.  The orbital stability
calculations of \citet{holman99} show that for a circularized binary
orbit (i.e., $e$~=~0) the critical radius for a system with a mass
ratio $>$~0.3 is $\sim$~2 times the orbital separation. Presuming the
debris disk to be circumbinary, we chose an inner radius of
1.3~\rsun\ and outer radii of 1.4, 1.5, 2.0, 3.0 and 5.0~\rsun. We
plot the result of our debris disk calculations in Figure
\ref{fg:disk}. If a debris disk corresponding to the parameters we
assumed for our models were present it would likely go
undetected. Only the widest and least inclined disks would be detected
in the $W1$ band, given the uncertainties. If J0745 does in fact
harbor a debris disk, it would have to be in the form of a narrow ring
or with an inclination angle high enough to limit the visible emission
from the disk. The $WISE$~data dictate that the disk could have an
inclination of $i \geq 30^{\circ}$ with $R_{\rm out} \leq
3.0$~\rsun. This would allow it to escape detection through IR
excess. Consequently, a debris disk scenario for J0745 cannot be
entirely ruled out.

\begin{figure}
\includegraphics[scale=0.45,bb=20 107 592 589]{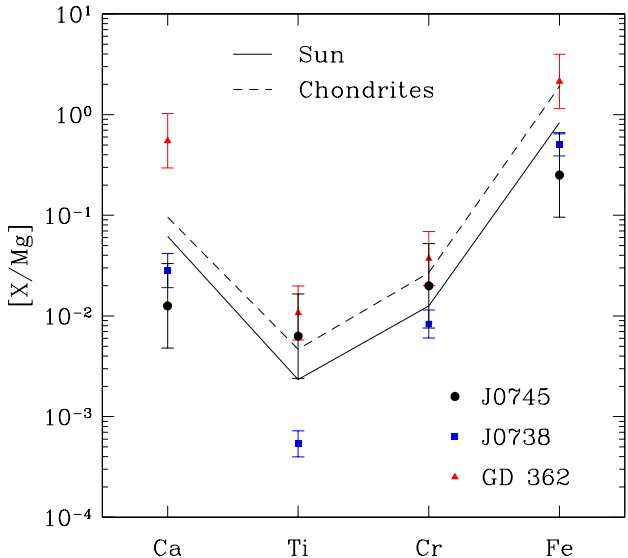}
\figcaption[f11.eps]{Elemental abundances by number relative to
  magnesium. The GD~362 data are from \citet{zuckerman07}, the J0738
  data are taken from \citet{dufour12}, and the solar and chondritic
  abundances are from \citet{lodders03}.  \label{fg:compZ}}
\end{figure}

To further explore the possibility that J0745 has accreted its metals
from a debris disk like other heavily polluted WDs, we compare the
abundance ratios of the detected metals with those seen in J0738 and
GD~362. Since we have not identified any Si in J0745, we use Mg as the
reference element and compute abundance ratios relative to Mg. The
results of this exercise are displayed in Figure \ref{fg:compZ}. We
also plot the abundance ratios for solar and chondritic material based
on the abundances in \citet{lodders03}. Unfortunately, the relatively
small number of metals detected in J0745 with our medium-resolution
MMT spectrum and the uncertainties in our abundance measurements
preclude any meaningful conclusions regarding the origin of metals in
J0745. Indeed, Figure \ref{fg:compZ} shows that all the abundance
ratios are generally similar with Ca and Fe being somewhat enhanced in
GD~362 while Ti is slightly deficient in J0738. We can also obtain a
rough estimate of the total mass of metals present in J0745. Assuming
a mass of $ M_{CZ} \approx 5 \times 10^{-9}$~\msun\ for the hydrogen
convection zone (L. Althaus 2013, private communication) coupled with
the metal abundances listed in Table \ref{tab2}, we estimate the total
mass of metals in the photosphere of J0745 to be $M_{Z} \gtrsim 5
\times 10^{22}$~g . This is roughly 1/20 the mass of the dwarf planet
Ceres, the largest asteroid in the solar system, and an order of
magnitude less than the mass of metals found in J0738
\citep{dufour12}. Our estimate necessarily represents a lower limit
since we have not yet identified all the metals present in the
atmosphere of J0745.

Another factor to consider is the diffusion timescale, $\tau_{D}$, for
metals in the atmosphere of an ELM WD. In more massive WDs, the
diffusion timescale is usually several orders of magnitude shorter
than the WD cooling age, $\tau_{\rm cool}$. This is one of the reasons
that accretion must necessarily be ongoing as any material from a
previous accretion episode would have long ago settled out of the
atmosphere. On the other hand, the much weaker surface gravity of ELM
WDs will likely allow heavy elements to diffuse more
slowly. \citet{koester06} compute diffusion times for normal WDs, but
not for the ELM WD regime. If we consider the models with
\Te~=~8000~K, we notice that the logarithm of the diffusion timescale
for Ca, Mg, and Fe, varies roughly linearly as a function of
\logg. Specifically, $\log \tau_{D}$ increases by $\sim$~0.5~dex for
every decrease of 0.25~dex in \logg\ for Mg and Fe, and by
$\sim$~0.4~dex for Ca. If we follow this trend, we can extrapolate
that for \logg~$\sim$~6.25, the values of $\log \tau_{D}$ for Ca, Mg
and Fe are $\sim$5.6, $\sim$6.7, and $\sim$6.6, respectively. The
models of \citet{althaus13} predict that the cooling age for J0745 is
$\log \tau_{\rm cool} \approx 9.6$. We see that even in ELM WDs, the
diffusion timescale is still much shorter than the evolutionary
timescale despite the weaker gravitational field. Note that the
calculations of \citet{koester06} assume a steady state accretion
scenario which may or may not be the case here. The accretion rate
depends both on the mass of the WD and the diffusion timescale. Both
of these quantities are not well constrained. Consequently, any
calculation of the accretion rate at this time would be speculative
and we leave such a determination for a future publication.

Yet another mechanism that might be at work in J0745 is radiative
levitation. It is well established that heavy elements can be
maintained in the atmospheres of very hot WDs by radiative forces
\citep[see e.g. ][]{chayer95a,chayer95b}. Recent calculations by
\citet{chayer10} demonstrate that WDs with \logg~=~8.0 and as cool as
\Te~=~20,000~K can potentially sustain C, Al and Si in their
atmospheres. J0745's surface gravity is two orders of magnitude weaker
and its temperature is 12,000~K cooler than the models presented in
\citet{chayer10}. Consequently, the radiation field might not be
intense enough, even at \logg~$\sim$~6.0, to overcome the local
gravitational force. However, radiative levitation calculations for
the appropriate regime of \Te\ and \logg\ are needed to conclusively
show if heavy elements can be sustained in the atmospheres of ELM WDs
by this mechanism.

It is also possible that a recent shell flash could have mixed up the
interior of the star bringing metals to the surface. A recent shell
flash is a plausible explanation for the WD companion to
PSR~J1816+4510 \citep{kaplan13} which has \Te~=~16,500 K and
\logg~=~4.90, and a cooling age of $\approx$~85~Myr
\citep{althaus13}. A recent shell flash seems unlikely for J0745 given
its mass (0.16~\msun) and cooling age (4.2~Gyr).  However, there is
still much uncertainty in ELM WD evolutionary models despite the
recent improvements presented in \citet{althaus13} when compared to
previously available calculations \citep[e.g.][]{panei07}. It would
therefore be premature to discount the shell flash scenario at this
point in time based solely on our determination of the cooling age.

\subsection{Surface Gravity}\label{gravity}

There is a discrepancy between the stellar radius measured from the
ellipsoidal variations and the stellar radius measured from our
\logg\ coupled with the \citet{althaus13} models. This is not the
first instance where we have encountered an issue in reconciling the
radius of an ELM WD, estimated through other means, with our
spectroscopic measurement of \logg. Indeed, the runaway binary
LP~400-22 poses an equally baffling mystery since its
\logg\ determined from Balmer line fitting is at odds with the
\logg\ inferred from parallax measurements \citep[for a detailed
  discussion, see][]{kilic13b}.  As a result, we must consider another
source of uncertainty. For normal mass WDs, a long standing issue has
been that of the ``high \logg\ problem" for WDs with \Te~$<$~13,000~K
\citep{kepler07,tremblay10,gianninas11}. Over the years, various
causes were proposed and explored. The root of the problem was finally
exposed by \citet{tremblay11} who showed that the 1D treatment of
convection using the mixing-length theory \citep[MLT;][]{bv58,bohm71}
was the true culprit. \citet{tremblay13a} presented new models that
employ a complete 3D hydrodynamical treatment of convection and
demonstrated that the new models effectively eliminate the so-called
``high \logg\ problem". If we consider that the atmosphere models we
rely on for our analysis use MLT to model convection, we should
logically expect them to suffer from the same high \logg\ problem as
well, albeit in a different regime of surface gravity. Recent
calculations show that for \logg~=~7.0, the 3D models predict a
surface gravity of $\sim$0.3 dex lower than the 1D models
\citep{tremblay13b}. Clearly, a grid of 3D models appropriate for ELM
WDs is needed to explore these effects further and to resolve the
discrepancy between the radius estimate from the photometric and
spectroscopic data.

\subsection{Future Evolution}

J0745 has a merger time less than a Hubble time (see Table
\ref{tab2}), it is of interest to explore what the ultimate fate of
this system will be. The mass ratio of the system is $q>0.36$. Binary
mass transfer models suggest that a system with this mass ratio will
undergo unstable mass transfer and merge to form a single WD
\citep{marsh04}.

\section{CONCLUSIONS}

We present the discovery of the most recent addition to the ELM WD
family, J0745. Spectroscopic and photometric analyses demonstrate that
J0745 is a cool, tidally distorted, compact ELM WD binary whose
primary is polluted with nearly solar abundances of Ca, Mg, Ti, Cr and
Fe. There is no evidence of IR excess from a debris disk from which
J0745 could be accreting heavy elements, as is often observed in more
massive WDs with atmospheric metal pollution. However, we cannot
completely rule out the debris disk scenario since the geometry of the
disk (inclination angle, radial extent) is unknown and could preclude
its detection at 2--5~$\mu$m. On the other hand, radiative levitation,
which would be favored in an ELM WD with its much weaker gravitational
field, may also play a role in keeping metals in the
atmosphere. Unfortunately, calculations for the radiative support of
heavy elements in the atmospheres of ELM WDs are not yet available.
Further progress requires a more accurate determination of the metal
abundances. The UV spectrum of J0745, accessible with the {\it Hubble
  Space Telescope's} {\it Cosmic Origins Spectrograph}, is
particularly useful for discriminating between the accretion,
radiative levitation, and shell flash scenarios.  \\ \\
\noindent We wish to thank the referee for a careful reading of this
manuscript and for useful suggestions that helped to improve it. We
also thank P. Canton for useful discussions. J.J.H. and
S.T.H. acknowledge the support of the NSF under grant AST-0909107 and
the Norman Hackerman Advanced Research Program under grant
003658-0252-2009. This research makes use of the SAO/NASA Astrophysics
Data System Bibliographic Service. This project makes use of data
products from the Sloan Digital Sky Survey, which has been funded by
the Alfred P. Sloan Foundation, the Participating Institutions, the
National Science Foundation, and the U.S. Department of Energy Office
of Science. M.K. acknowledges support from the NSF under grant
AST-1312678. This work was supported in part by the Smithsonian
Institution. This work is funded in part by the NSERC Canada.

{\it Facilities:} MMT (Blue Channel Spectrograph), FLWO 1.5m (FAST),
Struve (Argos)

\bibliographystyle{apj}
\bibliography{biblio}

\begin{thebibliography}{82}
\expandafter\ifx\csname natexlab\endcsname\relax\def\natexlab#1{#1}\fi

\bibitem[{{Ahn} {et~al.}(2012){Ahn}, {Alexandroff}, {Allende Prieto},
  {Anderson}, {Anderton}, {Andrews}, {Aubourg}, {Bailey}, {Balbinot}, {Barnes},
  \& et~al.}]{ahn12}
{Ahn}, C.~P., {Alexandroff}, R., {Allende Prieto}, C., {et~al.} 2012, \apjs,
  203, 21

\bibitem[{{Althaus} {et~al.}(2013){Althaus}, {Miller Bertolami}, \&
  {C{\'o}rsico}}]{althaus13}
{Althaus}, L.~G., {Miller Bertolami}, M.~M., \& {C{\'o}rsico}, A.~H. 2013,
  \aap, 557, A19

\bibitem[{{Asplund} {et~al.}(2005){Asplund}, {Grevesse}, \&
  {Sauval}}]{asplund05}
{Asplund}, M., {Grevesse}, N., \& {Sauval}, A.~J. 2005, in Astronomical Society
  of the Pacific Conference Series, Vol. 336, Cosmic Abundances as Records of
  Stellar Evolution and Nucleosynthesis, ed. T.~G. {Barnes}, III \& F.~N.
  {Bash}, 25

\bibitem[{{Barber} {et~al.}(2012){Barber}, {Patterson}, {Kilic}, {Leggett},
  {Dufour}, {Bloom}, \& {Starr}}]{barber12}
{Barber}, S.~D., {Patterson}, A.~J., {Kilic}, M., {et~al.} 2012, \apj, 760, 26

\bibitem[{{Berger} {et~al.}(2005){Berger}, {Koester}, {Napiwotzki}, {Reid}, \&
  {Zuckerman}}]{berger05}
{Berger}, L., {Koester}, D., {Napiwotzki}, R., {Reid}, I.~N., \& {Zuckerman},
  B. 2005, \aap, 444, 565

\bibitem[{{Bildsten} {et~al.}(2007){Bildsten}, {Shen}, {Weinberg}, \&
  {Nelemans}}]{bildsten07}
{Bildsten}, L., {Shen}, K.~J., {Weinberg}, N.~N., \& {Nelemans}, G. 2007,
  \apjl, 662, L95

\bibitem[{{Bloemen} {et~al.}(2012){Bloemen}, {Marsh}, {Degroote},
  {{\O}stensen}, {P{\'a}pics}, {Aerts}, {Koester}, {G{\"a}nsicke}, {Breedt},
  {Lombaert}, {Pyrzas}, {Copperwheat}, {Exter}, {Raskin}, {Van Winckel},
  {Prins}, {Pessemier}, {Fr{\'e}mat}, {Hensberge}, {Jorissen}, \& {Van
  Eck}}]{bloemen12}
{Bloemen}, S., {Marsh}, T.~R., {Degroote}, P., {et~al.} 2012, \mnras, 422, 2600

\bibitem[{{Bohm} \& {Cassinelli}(1971)}]{bohm71}
{Bohm}, K.~H., \& {Cassinelli}, J. 1971, \aap, 12, 21

\bibitem[{{B{\"o}hm-Vitense}(1958)}]{bv58}
{B{\"o}hm-Vitense}, E. 1958, \zap, 46, 108

\bibitem[{{Brown} {et~al.}(2013){Brown}, {Kilic}, {Allende Prieto},
  {Gianninas}, \& {Kenyon}}]{brown_ELM5}
{Brown}, W.~R., {Kilic}, M., {Allende Prieto}, C., {Gianninas}, A., \&
  {Kenyon}, S.~J. 2013, \apj, 769, 66

\bibitem[{{Brown} {et~al.}(2010){Brown}, {Kilic}, {Allende Prieto}, \&
  {Kenyon}}]{brown_ELM1}
{Brown}, W.~R., {Kilic}, M., {Allende Prieto}, C., \& {Kenyon}, S.~J. 2010,
  \apj, 723, 1072

\bibitem[{{Brown} {et~al.}(2012){Brown}, {Kilic}, {Allende Prieto}, \&
  {Kenyon}}]{brown_ELM3}
---. 2012, \apj, 744, 142

\bibitem[{{Chayer}(2013)}]{chayer13}
{Chayer}, P. 2013, arXiv:1310.6245

\bibitem[{{Chayer} \& {Dupuis}(2010)}]{chayer10}
{Chayer}, P., \& {Dupuis}, J. 2010, in American Institute of Physics Conference
  Series, Vol. 1273, American Institute of Physics Conference Series, ed.
  K.~{Werner} \& T.~{Rauch}, 394--399

\bibitem[{{Chayer} {et~al.}(1995{\natexlab{a}}){Chayer}, {Fontaine}, \&
  {Wesemael}}]{chayer95a}
{Chayer}, P., {Fontaine}, G., \& {Wesemael}, F. 1995{\natexlab{a}}, \apjs, 99,
  189

\bibitem[{{Chayer} {et~al.}(1995{\natexlab{b}}){Chayer}, {Vennes}, {Pradhan},
  {Thejll}, {Beauchamp}, {Fontaine}, \& {Wesemael}}]{chayer95b}
{Chayer}, P., {Vennes}, S., {Pradhan}, A.~K., {et~al.} 1995{\natexlab{b}},
  \apj, 454, 429

\bibitem[{{Debes} {et~al.}(2012){Debes}, {Hoard}, {Farihi}, {Wachter},
  {Leisawitz}, \& {Cohen}}]{debes12}
{Debes}, J.~H., {Hoard}, D.~W., {Farihi}, J., {et~al.} 2012, \apj, 759, 37

\bibitem[{{Debes} \& {Sigurdsson}(2002)}]{debes02}
{Debes}, J.~H., \& {Sigurdsson}, S. 2002, \apj, 572, 556

\bibitem[{{Dufour} {et~al.}(2010){Dufour}, {Kilic}, {Fontaine}, {Bergeron},
  {Lachapelle}, {Kleinman}, \& {Leggett}}]{dufour10}
{Dufour}, P., {Kilic}, M., {Fontaine}, G., {et~al.} 2010, \apj, 719, 803

\bibitem[{{Dufour} {et~al.}(2012){Dufour}, {Kilic}, {Fontaine}, {Bergeron},
  {Melis}, \& {Bochanski}}]{dufour12}
---. 2012, \apj, 749, 6

\bibitem[{{Fabricant} {et~al.}(1998){Fabricant}, {Cheimets}, {Caldwell}, \&
  {Geary}}]{fabricant98}
{Fabricant}, D., {Cheimets}, P., {Caldwell}, N., \& {Geary}, J. 1998, \pasp,
  110, 79

\bibitem[{{Farihi}(2011)}]{farihi11}
{Farihi}, J. 2011, in American Institute of Physics Conference Series, Vol.
  1331, American Institute of Physics Conference Series, ed. S.~{Schuh},
  H.~{Drechsel}, \& U.~{Heber}, 193--210

\bibitem[{{Farihi} {et~al.}(2010{\natexlab{a}}){Farihi}, {Barstow}, {Redfield},
  {Dufour}, \& {Hambly}}]{farihi10a}
{Farihi}, J., {Barstow}, M.~A., {Redfield}, S., {Dufour}, P., \& {Hambly},
  N.~C. 2010{\natexlab{a}}, \mnras, 404, 2123

\bibitem[{{Farihi} {et~al.}(2010{\natexlab{b}}){Farihi}, {Jura}, {Lee}, \&
  {Zuckerman}}]{farihi10b}
{Farihi}, J., {Jura}, M., {Lee}, J.-E., \& {Zuckerman}, B. 2010{\natexlab{b}},
  \apj, 714, 1386

\bibitem[{{Farihi} {et~al.}(2009){Farihi}, {Jura}, \& {Zuckerman}}]{farihi09}
{Farihi}, J., {Jura}, M., \& {Zuckerman}, B. 2009, \apj, 694, 805

\bibitem[{{Gianninas} {et~al.}(2011){Gianninas}, {Bergeron}, \&
  {Ruiz}}]{gianninas11}
{Gianninas}, A., {Bergeron}, P., \& {Ruiz}, M.~T. 2011, \apj, 743, 138

\bibitem[{{Gianninas} {et~al.}(2004){Gianninas}, {Dufour}, \&
  {Bergeron}}]{gianninas04}
{Gianninas}, A., {Dufour}, P., \& {Bergeron}, P. 2004, \apjl, 617, L57

\bibitem[{{Gianninas} {et~al.}(2013){Gianninas}, {Strickland}, {Kilic}, \&
  {Bergeron}}]{gianninas13}
{Gianninas}, A., {Strickland}, B.~D., {Kilic}, M., \& {Bergeron}, P. 2013,
  \apj, 766, 3

\bibitem[{{Hermes} {et~al.}(2012{\natexlab{a}}){Hermes}, {Kilic}, {Brown},
  {Montgomery}, \& {Winget}}]{hermes12a}
{Hermes}, J.~J., {Kilic}, M., {Brown}, W.~R., {Montgomery}, M.~H., \& {Winget},
  D.~E. 2012{\natexlab{a}}, \apj, 749, 42

\bibitem[{{Hermes} {et~al.}(2012{\natexlab{b}}){Hermes}, {Kilic}, {Brown},
  {Winget}, {Allende Prieto}, {Gianninas}, {Mukadam}, {Cabrera-Lavers}, \&
  {Kenyon}}]{hermes12b}
{Hermes}, J.~J., {Kilic}, M., {Brown}, W.~R., {et~al.} 2012{\natexlab{b}},
  \apjl, 757, L21

\bibitem[{{Hermes} {et~al.}(2013){Hermes}, {Montgomery}, {Winget}, {Brown},
  {Gianninas}, {Kilic}, {Kenyon}, {Bell}, \& {Harrold}}]{hermes13}
{Hermes}, J.~J., {Montgomery}, M.~H., {Winget}, D.~E., {et~al.} 2013, \apj,
  765, 102

\bibitem[{{Hoard} {et~al.}(2013){Hoard}, {Debes}, {Wachter}, {Leisawitz}, \&
  {Cohen}}]{hoard13}
{Hoard}, D.~W., {Debes}, J.~H., {Wachter}, S., {Leisawitz}, D.~T., \& {Cohen},
  M. 2013, \apj, 770, 21

\bibitem[{{Holberg} \& {Bergeron}(2006)}]{holberg06}
{Holberg}, J.~B., \& {Bergeron}, P. 2006, \aj, 132, 1221

\bibitem[{{Holman} \& {Wiegert}(1999)}]{holman99}
{Holman}, M.~J., \& {Wiegert}, P.~A. 1999, \aj, 117, 621

\bibitem[{{Hummer} \& {Mihalas}(1988)}]{hm88}
{Hummer}, D.~G., \& {Mihalas}, D. 1988, \apj, 331, 794

\bibitem[{{Jura}(2003)}]{jura03}
{Jura}, M. 2003, \apjl, 584, L91

\bibitem[{{Jura}(2006)}]{jura06}
---. 2006, \apj, 653, 613

\bibitem[{{Jura}(2008)}]{jura08}
---. 2008, \aj, 135, 1785

\bibitem[{{Jura} {et~al.}(2007){Jura}, {Farihi}, \& {Zuckerman}}]{jura07}
{Jura}, M., {Farihi}, J., \& {Zuckerman}, B. 2007, \apj, 663, 1285

\bibitem[{{Kanaan} {et~al.}(2002){Kanaan}, {Kepler}, \& {Winget}}]{kanaan02}
{Kanaan}, A., {Kepler}, S.~O., \& {Winget}, D.~E. 2002, \aap, 389, 896

\bibitem[{{Kaplan} {et~al.}(2013){Kaplan}, {Bhalerao}, {van Kerkwijk},
  {Koester}, {Kulkarni}, \& {Stovall}}]{kaplan13}
{Kaplan}, D.~L., {Bhalerao}, V.~B., {van Kerkwijk}, M.~H., {et~al.} 2013, \apj,
  765, 158

\bibitem[{{Kenyon} \& {Garcia}(1986)}]{kenyon86}
{Kenyon}, S.~J., \& {Garcia}, M.~R. 1986, \aj, 91, 125

\bibitem[{{Kepler} {et~al.}(2007){Kepler}, {Kleinman}, {Nitta}, {Koester},
  {Castanheira}, {Giovannini}, {Costa}, \& {Althaus}}]{kepler07}
{Kepler}, S.~O., {Kleinman}, S.~J., {Nitta}, A., {et~al.} 2007, \mnras, 375,
  1315

\bibitem[{{Kilic} {et~al.}(2011){Kilic}, {Brown}, {Allende Prieto},
  {Ag{\"u}eros}, {Heinke}, \& {Kenyon}}]{kilic_ELM2}
{Kilic}, M., {Brown}, W.~R., {Allende Prieto}, C., {et~al.} 2011, \apj, 727, 3

\bibitem[{{Kilic} {et~al.}(2012{\natexlab{a}}){Kilic}, {Brown}, {Allende
  Prieto}, {Kenyon}, {Heinke}, {Ag{\"u}eros}, \& {Kleinman}}]{kilic_ELM4}
---. 2012{\natexlab{a}}, \apj, 751, 141

\bibitem[{{Kilic} {et~al.}(2013{\natexlab{a}}){Kilic}, {Brown}, \&
  {Hermes}}]{kilic13a}
{Kilic}, M., {Brown}, W.~R., \& {Hermes}, J.~J. 2013{\natexlab{a}}, in
  Astronomical Society of the Pacific Conference Series, Vol. 467, 9th LISA
  Symposium, ed. G.~{Auger}, P.~{Bin{\'e}truy}, \& E.~{Plagnol}, 47

\bibitem[{{Kilic} {et~al.}(2012{\natexlab{b}}){Kilic}, {Patterson}, {Barber},
  {Leggett}, \& {Dufour}}]{kilic12}
{Kilic}, M., {Patterson}, A.~J., {Barber}, S., {Leggett}, S.~K., \& {Dufour},
  P. 2012{\natexlab{b}}, \mnras, 419, L59

\bibitem[{{Kilic} {et~al.}(2006){Kilic}, {von Hippel}, {Leggett}, \&
  {Winget}}]{kilic06}
{Kilic}, M., {von Hippel}, T., {Leggett}, S.~K., \& {Winget}, D.~E. 2006, \apj,
  646, 474

\bibitem[{{Kilic} {et~al.}(2013{\natexlab{b}}){Kilic}, {Hermes}, {Gianninas},
  {Brown}, {Heinke}, {Agueros}, {Chote}, {Sullivan}, {Bell}, \&
  {Harrold}}]{kilic13c}
{Kilic}, M., {Hermes}, J.~J., {Gianninas}, A., {et~al.} 2013{\natexlab{b}},
  arXiv:1310.6359

\bibitem[{{Kilic} {et~al.}(2013{\natexlab{c}}){Kilic}, {Gianninas}, {Brown},
  {Harris}, {Dahn}, {Ag{\"u}eros}, {Heinke}, {Kenyon}, {Panei}, \&
  {Camilo}}]{kilic13b}
{Kilic}, M., {Gianninas}, A., {Brown}, W.~R., {et~al.} 2013{\natexlab{c}},
  \mnras, 434, 3582

\bibitem[{{Klein} {et~al.}(2011){Klein}, {Jura}, {Koester}, \&
  {Zuckerman}}]{klein11}
{Klein}, B., {Jura}, M., {Koester}, D., \& {Zuckerman}, B. 2011, \apj, 741, 64

\bibitem[{{Klein} {et~al.}(2010){Klein}, {Jura}, {Koester}, {Zuckerman}, \&
  {Melis}}]{klein10}
{Klein}, B., {Jura}, M., {Koester}, D., {Zuckerman}, B., \& {Melis}, C. 2010,
  \apj, 709, 950

\bibitem[{{Koester} {et~al.}(2005){Koester}, {Rollenhagen}, {Napiwotzki},
  {Voss}, {Christlieb}, {Homeier}, \& {Reimers}}]{koester05}
{Koester}, D., {Rollenhagen}, K., {Napiwotzki}, R., {et~al.} 2005, \aap, 432,
  1025

\bibitem[{{Koester} \& {Wilken}(2006)}]{koester06}
{Koester}, D., \& {Wilken}, D. 2006, \aap, 453, 1051

\bibitem[{{Liebert} {et~al.}(2005){Liebert}, {Bergeron}, \& {Holberg}}]{LBH05}
{Liebert}, J., {Bergeron}, P., \& {Holberg}, J.~B. 2005, \apjs, 156, 47

\bibitem[{{Lodders}(2003)}]{lodders03}
{Lodders}, K. 2003, \apj, 591, 1220

\bibitem[{{Marsh} {et~al.}(2004){Marsh}, {Nelemans}, \& {Steeghs}}]{marsh04}
{Marsh}, T.~R., {Nelemans}, G., \& {Steeghs}, D. 2004, \mnras, 350, 113

\bibitem[{{Martin} {et~al.}(2005){Martin}, {Fanson}, {Schiminovich},
  {Morrissey}, {Friedman}, {Barlow}, {Conrow}, {Grange}, {Jelinsky},
  {Milliard}, {Siegmund}, {Bianchi}, {Byun}, {Donas}, {Forster}, {Heckman},
  {Lee}, {Madore}, {Malina}, {Neff}, {Rich}, {Small}, {Surber}, {Szalay},
  {Welsh}, \& {Wyder}}]{martin05}
{Martin}, D.~C., {Fanson}, J., {Schiminovich}, D., {et~al.} 2005, \apjl, 619,
  L1

\bibitem[{{Melis} {et~al.}(2010){Melis}, {Jura}, {Albert}, {Klein}, \&
  {Zuckerman}}]{melis10}
{Melis}, C., {Jura}, M., {Albert}, L., {Klein}, B., \& {Zuckerman}, B. 2010,
  \apj, 722, 1078

\bibitem[{{Morris} \& {Naftilan}(1993)}]{morris93}
{Morris}, S.~L., \& {Naftilan}, S.~A. 1993, \apj, 419, 344

\bibitem[{{Nather} \& {Mukadam}(2004)}]{nather04}
{Nather}, R.~E., \& {Mukadam}, A.~S. 2004, \apj, 605, 846

\bibitem[{{Panei} {et~al.}(2007){Panei}, {Althaus}, {Chen}, \& {Han}}]{panei07}
{Panei}, J.~A., {Althaus}, L.~G., {Chen}, X., \& {Han}, Z. 2007, \mnras, 382,
  779

\bibitem[{{Perryman} {et~al.}(2001){Perryman}, {de Boer}, {Gilmore}, {H{\o}g},
  {Lattanzi}, {Lindegren}, {Luri}, {Mignard}, {Pace}, \& {de
  Zeeuw}}]{perryman01}
{Perryman}, M.~A.~C., {de Boer}, K.~S., {Gilmore}, G., {et~al.} 2001, \aap,
  369, 339

\bibitem[{{Rafikov}(2011{\natexlab{a}})}]{rafikov11a}
{Rafikov}, R.~R. 2011{\natexlab{a}}, \apjl, 732, L3

\bibitem[{{Rafikov}(2011{\natexlab{b}})}]{rafikov11b}
---. 2011{\natexlab{b}}, \mnras, 416, L55

\bibitem[{{Rafikov} \& {Garmilla}(2012)}]{rafikov12}
{Rafikov}, R.~R., \& {Garmilla}, J.~A. 2012, \apj, 760, 123

\bibitem[{{Schlegel} {et~al.}(1998){Schlegel}, {Finkbeiner}, \&
  {Davis}}]{schlegel98}
{Schlegel}, D.~J., {Finkbeiner}, D.~P., \& {Davis}, M. 1998, \apj, 500, 525

\bibitem[{{Schmidt} {et~al.}(1989){Schmidt}, {Weymann}, \& {Foltz}}]{schmidt89}
{Schmidt}, G.~D., {Weymann}, R.~J., \& {Foltz}, C.~B. 1989, \pasp, 101, 713

\bibitem[{{Shporer} {et~al.}(2010){Shporer}, {Kaplan}, {Steinfadt}, {Bildsten},
  {Howell}, \& {Mazeh}}]{shporer10}
{Shporer}, A., {Kaplan}, D.~L., {Steinfadt}, J.~D.~R., {et~al.} 2010, \apjl,
  725, L200

\bibitem[{{Skrutskie} {et~al.}(2006){Skrutskie}, {Cutri}, {Stiening},
  {Weinberg}, {Schneider}, {Carpenter}, {Beichman}, {Capps}, {Chester},
  {Elias}, {Huchra}, {Liebert}, {Lonsdale}, {Monet}, {Price}, {Seitzer},
  {Jarrett}, {Kirkpatrick}, {Gizis}, {Howard}, {Evans}, {Fowler}, {Fullmer},
  {Hurt}, {Light}, {Kopan}, {Marsh}, {McCallon}, {Tam}, {Van Dyk}, \&
  {Wheelock}}]{skrutskie06}
{Skrutskie}, M.~F., {Cutri}, R.~M., {Stiening}, R., {et~al.} 2006, \aj, 131,
  1163

\bibitem[{{Stumpff}(1980)}]{stumpff80}
{Stumpff}, P. 1980, \aaps, 41, 1

\bibitem[{{Thompson} \& {Mullally}(2009)}]{thompson09}
{Thompson}, S.~E., \& {Mullally}, F. 2009, Journal of Physics Conference
  Series, 172, 012081

\bibitem[{{Tremblay} \& {Bergeron}(2009)}]{TB09}
{Tremblay}, P.-E., \& {Bergeron}, P. 2009, \apj, 696, 1755

\bibitem[{{Tremblay} {et~al.}(2010){Tremblay}, {Bergeron}, {Kalirai}, \&
  {Gianninas}}]{tremblay10}
{Tremblay}, P.-E., {Bergeron}, P., {Kalirai}, J.~S., \& {Gianninas}, A. 2010,
  \apj, 712, 1345

\bibitem[{{Tremblay} {et~al.}(2011){Tremblay}, {Ludwig}, {Steffen}, {Bergeron},
  \& {Freytag}}]{tremblay11}
{Tremblay}, P.-E., {Ludwig}, H.-G., {Steffen}, M., {Bergeron}, P., \&
  {Freytag}, B. 2011, \aap, 531, L19

\bibitem[{{Tremblay} {et~al.}(2013{\natexlab{a}}){Tremblay}, {Ludwig},
  {Steffen}, \& {Freytag}}]{tremblay13a}
{Tremblay}, P.-E., {Ludwig}, H.-G., {Steffen}, M., \& {Freytag}, B.
  2013{\natexlab{a}}, \aap, 552, A13

\bibitem[{{Tremblay} {et~al.}(2013{\natexlab{b}}){Tremblay}, {Ludwig},
  {Steffen}, \& {Freytag}}]{tremblay13b}
---. 2013{\natexlab{b}}, arXiv:1309.0886

\bibitem[{{von Hippel} {et~al.}(2007){von Hippel}, {Kuchner}, {Kilic},
  {Mullally}, \& {Reach}}]{vh07}
{von Hippel}, T., {Kuchner}, M.~J., {Kilic}, M., {Mullally}, F., \& {Reach},
  W.~T. 2007, \apj, 662, 544

\bibitem[{{Wright} {et~al.}(2010){Wright}, {Eisenhardt}, {Mainzer}, {Ressler},
  {Cutri}, {Jarrett}, {Kirkpatrick}, {Padgett}, {McMillan}, {Skrutskie},
  {Stanford}, {Cohen}, {Walker}, {Mather}, {Leisawitz}, {Gautier}, {McLean},
  {Benford}, {Lonsdale}, {Blain}, {Mendez}, {Irace}, {Duval}, {Liu}, {Royer},
  {Heinrichsen}, {Howard}, {Shannon}, {Kendall}, {Walsh}, {Larsen}, {Cardon},
  {Schick}, {Schwalm}, {Abid}, {Fabinsky}, {Naes}, \& {Tsai}}]{wright10}
{Wright}, E.~L., {Eisenhardt}, P.~R.~M., {Mainzer}, A.~K., {et~al.} 2010, \aj,
  140, 1868

\bibitem[{{Wyder} {et~al.}(2007){Wyder}, {Martin}, {Schiminovich}, {Seibert},
  {Budav{\'a}ri}, {Treyer}, {Barlow}, {Forster}, {Friedman}, {Morrissey},
  {Neff}, {Small}, {Bianchi}, {Donas}, {Heckman}, {Lee}, {Madore}, {Milliard},
  {Rich}, {Szalay}, {Welsh}, \& {Yi}}]{wyder07}
{Wyder}, T.~K., {Martin}, D.~C., {Schiminovich}, D., {et~al.} 2007, \apjs, 173,
  293

\bibitem[{{Zuckerman} {et~al.}(2007){Zuckerman}, {Koester}, {Melis}, {Hansen},
  \& {Jura}}]{zuckerman07}
{Zuckerman}, B., {Koester}, D., {Melis}, C., {Hansen}, B.~M., \& {Jura}, M.
  2007, \apj, 671, 872

\bibitem[{{Zuckerman} {et~al.}(2003){Zuckerman}, {Koester}, {Reid}, \&
  {H{\"u}nsch}}]{zuckerman03}
{Zuckerman}, B., {Koester}, D., {Reid}, I.~N., \& {H{\"u}nsch}, M. 2003, \apj,
  596, 477

\end{thebibliography}

\end{document}